\documentclass[useAMS]{mn2e}
\usepackage{graphicx,times}

\def\eq#1{\begin{equation} #1 \end{equation}}

\def\DS     {\displaystyle}
\def\Ref    {\item}
\def\E#1{\hbox{$10^{#1}$}}
\def\sub#1{_{\rm #1}}
\def\case#1/#2{\hbox{$\frac{#1}{#2}$}}
\def\about  {\hbox{$\sim$}}
\def\x      {\hbox{$\times$}}
\def\PsiO   {\hbox{$\Psi_0$}}
\def\Fbol   {\hbox{$F\sub{bol}$}}

\def\QV     {\hbox{$Q\sub V$}}
\def\Qast   {\hbox{$Q_\ast$}}
\def\tV     {\hbox{$\tau\sub V$}}
\def\tF     {\hbox{$\tau\sub F$}}
\def\vf     {\hbox{$v_\infty$}}

\def\ss     {\hbox{$\sigma_{22}$}}

\def\kms    {\hbox{km s$^{-1}$}}

\def\nd     {\hbox{$n\sub d$}}
\def\nH     {\hbox{$n\sub H$}}

\def\Mdot   {\hbox{$\dot M$}}
\def\Msix   {\hbox{$\dot M_{-6}$}}

\def\Lo     {\hbox{$L_\odot$}}
\def\Mo     {\hbox{$M_\odot$}}
\def\Myr    {\hbox{\Mo\ yr$^{-1}$}}
\def\mic    {\hbox{$\umu$m}}

\def\Tc     {\hbox{$T\sub{c}$}}

\def\rdg    {\hbox{$r\sub{dg}$}}

\def\xi     {\hbox{$x_{1i}$}}

\title[Dusty Winds II]
               {Dusty winds --- II. Observational Implications}

\author[\v{Z}. Ivezi\'c and M. Elitzur]
       {\v Zeljko Ivezi\'c$^1$ and Moshe Elitzur$^2$ \\
   $^1$Department of Astronomy,
       University of Washington,
       Box 351580,
       Seattle, WA 98195,
       USA;
       ivezic@astro.washington.edu\\
   $^2$Department of Physics and Astronomy,
       University of Kentucky,
       Lexington, KY 40506-0055,
       USA;
       moshe@pa.uky.edu\\}

\date{Submitted Dec 3, 2009}

\pagerange{\pageref{firstpage}--\pageref{lastpage}} \pubyear{2010}

\begin{document}
\label{firstpage}

\maketitle

\begin{abstract}
We compare observations of AGB stars and predictions of the Elitzur \&
Ivezi\'{c} (2001) steady-state radiatively driven dusty wind model. The model
results are described by a set of similarity functions of a single independent
variable, and imply general scaling relations among the system parameters. We
find that the model properly reproduces various correlations among the observed
quantities and demonstrate that dust drift through the gas has a major impact
on the structure of most winds. From data for nearby oxygen-rich and
carbon-rich mass-losing stars we find that (1) the dispersion in grain
properties within each group is rather small; (2) both the dust cross-section
per gas particle and the dust-to-gas mass ratio are similar for the two samples
even though the stellar atmospheres and grain properties are very different;
(3) the dust abundance in both outflows is significantly below the Galactic
average, indicating that most of the Galactic dust is not stardust---contrary
to popular belief, but in support of Draine (2009). Our model results can be
easily applied to recent massive data sets, such as the Spitzer SAGE survey of
the Large Magellanic Cloud, and incorporated in galaxy evolution models.

\end{abstract}


\section{                         INTRODUCTION                               }

Winds blown by stars on the asymptotic giant branch (AGB) are an important
component of mass return into the interstellar medium and may account for a
significant fraction of interstellar dust, including dust formation in
the early universe (Sloan et al. 2009). Therefore in addition to its
obvious significance for the theory of stellar evolution, the study of
AGB winds has important implications for the structure and evolution of
galaxies (Girardi \& Marigo 2007; Marston et al. 2009). In our own
Galaxy, its estimated 200,000 AGB stars are a good tracer of dominant
components, including the bulge (Whitelock \& Feast 2000; Jackson,
Ivezi\'{c} \& Knapp 2002). AGB stars also have a great potential as
distance indicators (Rejkuba 2004).

Over the past few years, data of AGB stars have improved significantly. The
2MASS survey delivered an all-sky near-IR photometric catalog to $K<14$, and
Spitzer obtained mid-IR (IRAC and MIPS) photometry for several thousand AGB
stars from the Large Magellanic Cloud (LMC, Meixner et al. 2006). In addition,
the MACHO survey provided high-quality light curves for $\sim$22,000 AGB stars
from the LMC (Fraser et al. 2005), and the Northern Sky Variability Survey
provided light curves for another $\sim$9,000 AGB stars brighter than
$V\sim15.5$ and with declination $\delta>-38^\circ$ (Wo\'{z}niak et al. 2004).
These new accurate and massive data sets are expected to rejuvenate studies of
AGB stars, leading to model-based interpretation of diverse measurements
including photometry, outflow velocity, mass-loss rate, pulsations, etc. (e.g.,
Marigo et al. 2008).

AGB stars are surrounded by an expanding envelope composed of gas and dust,
that has a major impact on observed properties. The complete description of
such a dusty wind should start with a full dynamic atmosphere model and
incorporate the processes that initiate the outflow and set the value of \Mdot.
These processes are yet to be identified with certainty, the most promising are
stellar pulsation (e.g. Bowen 1989; van Loon et al. 2008) and radiation
pressure on the water molecules (e.g. Elitzur, Brown \& Johnson 1989). Proper
description of these processes should be followed by grain formation and
growth, and subsequent wind dynamics. Two ambitious programs attempting to
incorporate as many aspects of this formidable task as possible have been
conducted over the last decade by groups at Berlin (see Winters et al. 2000;
Wachter et al. 2008; and references therein) and Vienna (see H\"{o}fner 1999;
Dorfi et al. 2001; H\"{o}fner et al. 2003; Nowotny et al. 2005; and references
therein). While much has been accomplished, the complexity of this undertaking
necessitates simplifications such as a pulsating boundary. In spite of
continuous progress, detailed understanding of atmospheric dynamics and grain
formation is still far from complete (e.g., H\"{o}fner \& Andersen 2007).

Fortunately, the full problem splits naturally to two parts, as recognized long
ago by Goldreich \& Scoville (1976). Once radiation pressure on the dust
exceeds all other forces, the rapid acceleration to supersonic velocities
decouples the outflow from the earlier phases---{\em the supersonic phase would
be exactly the same in two different outflows if they had the same mass-loss
rate and grain properties even if the grains were produced by entirely
different processes.} Furthermore, these stages are controlled by processes
that are much less dependent on detailed micro-physics, and are reasonably well
understood. And since most observations probe only the supersonic phase, models
devoted exclusively to this stage should reproduce the observable results while
avoiding the pitfalls and uncertainties of dust formation and the wind
initiation.

In Elitzur \& Ivezi\'{c} (2001, hereafter paper I) we present the
self-similarity solution of the dusty wind steady state supersonic phase. The
model assumes steady-state spherically symmetric mass loss with prompt dust
formation and no subsequent change in dust properties (i.e., no grain growth
and sputtering). The included dynamical effects are the radiation pressure
force, dust drift through the gas, and the gravitational pull by the star.
Paper I contains the description of the model and its solution in full
mathematical rigor. Here we discuss observational implications of the model,
and compare them to available data. The main questions that we ask are
\begin{itemize}
\item Can the model explain the wind velocity profile, including
         the region of strong acceleration at small radii?
\item Does the implied density profile produce spectral energy
         distributions in agreement with observations?
\item Do dynamical quantities, such as the wind terminal velocity and
         mass-loss rate, and photometric quantities such as colors and
          bolometric flux, display correlations predicted by the model?
\item How similar, or dissimilar, are dust properties inferred from
      observations of oxygen-rich and carbon-rich AGB stars?
\end{itemize}

We first discuss dynamical quantities, and then focus on the spectral energy
distribution. In \S2 we describe the wind velocity profile. Section 3 presents
correlations among final velocity \vf, mass loss rate \Mdot\ and luminosity
$L$. In \S4 we discuss the spectral energy distribution (SED) and its
relationship with the wind dynamical properties. In \S5 we discuss the allowed
range of physical parameters of dusty winds. The results are summarized and
discussed in \S6. Equations from paper I are referred to by their number
preceded by I.

\section{Velocity Profile}

The velocity profile for all winds, whether the dust is comprised of carbon or
silicate grains, can be summarized with the simple analytic expression (see \S3
in paper I)
\eq{
\label{eq:velprof}
    v = \vf\left(1 - {\theta_0\over\theta}\right)^k,     \qquad \hbox{where}\
  \cases{k    =   \frac23  & when $\tV \la 1$ \cr
         k \simeq    0.4   & when $\tV \ga 1$ \cr}
}
Here $\theta$ is angular distance from the star in the plane of the sky and
$\theta_0$ is a characteristic angular scale. The power $k$ introduces a weak
dependence on \tV, the overall radial optical depth at visual. The small-\tV\
value $k = \frac23$ reflects the effect of the drift, which dominates the
dynamics in that regime. At large \tV, reddening effects a switch to a more
moderate profile with $k \simeq 0.4$.

We discuss the wind terminal velocity \vf\ in \S3 below. Here we concentrate on
the shape of the radial profile $v/\vf$. The profile is mainly controlled by
the characteristic angular scale $\theta_0$, corresponding to the dust
condensation radius. It can be determined from the measured bolometric flux,
$\Fbol$, via
\eq{\label{eq:theta0}
   \theta_0 = 0.04\,\Psi^{1/2} \left({1000\,{\rm K} \over \Tc}\right)^{\!1\!/2}
                \left(\Fbol \over 10^{-8}\,{\rm W\,m^{-2}}\right)^{\!1\!/2}
                \hbox{ arcsec}.
}
Here \Tc\ is the dust condensation temperature and $\Psi$ is a dimensionless
function that sets the dust condensation radius as $\frac12\,r_\ast\Psi^{1/2}
(T_\ast/\Tc)^2$, where $r_\ast$ and $T_\ast$ are, respectively, the stellar
radius and effective temperature.  All our model calculations assume \Tc\ = 700
K and a black-body star with $T_\ast$ = 2500 K. The function $\Psi$ is
determined from the radiative transfer solution (see I.41 and I.42) and can be
adequately approximated by the simple analytic expression
\eq{
\label{eq:psi}
    \Psi = \Psi_0(1 + 0.005\tau_{\rm V}^m).
}
The parameters $\Psi_0$ and $m$ are listed in Table~\ref{tab:dust} together
with other relevant properties of the dust grains. In all the numerical
calculations here we employ amorphous carbon grains with optical properties
from Hanner (1988), and (the ``warm'' version of) silicate grains from
Ossenkopf, Henning \& Mathis (1992). We assume a single size $a$ = 0.1 \mic\ to
compute the grain absorption and scattering efficiencies; the grain size does
not enter the solution of the dusty wind problem in any other way. From their
effect on the variation of angular scale $\theta_0$ with \tV, the parameters
$\Psi_0$ and $m$ introduce the only dependence of the velocity profile on dust
properties. However, in practice this effect is rather weak: $\theta_0$ varies
by less than 10\% for \tV\ between 0 and 20.
%%%%%%%%%%%%%%%%%%%%%%%%%%%%% Table 1 %%%%%%%%%%%%%%%%%%%%%%%%%%%%%
\begin{table}
\begin{center}
\begin{tabular}{lrr}
\hline
               &  Carbon & Silicate  \\
\hline \hline
   \QV         &  2.40   &  1.15  \\
   \Qast       &  0.60   &  0.11  \\
   \PsiO       &  5.97   &  2.72  \\
     $m$       &   1.0   &  1.25  \\
\hline
\end{tabular}

\caption{Standard parameters for dust grains with size $a$ = 0.1\mic: \QV\ is
the efficiency factor for absorption at visual; \Qast\ is the Planck average at
the stellar temperature (2,500\,K) of the efficiency coefficient for radiation
pressure (see paper I, equation 4); \PsiO\ and $m$, defined in
eq.~\ref{eq:psi}, determine the quantity $\Psi$ that enters into the expression
for the dust sublimation angular size (eq.\ \ref{eq:theta0}).
\label{tab:dust}}
\end{center}
\end{table}
%%%%%%%%%%%%%%%%%%%%%%%%%%%%%%%%%%%%%%%%%%%%%%%%%%%%%%%%%%%%%%%%%%%

%%%%%%%%%%%%%%%%%%%% %%% %%%%%%%%%%%%%%%%%%%%%%%%%%%%%%%%%%%%%%%%%%%%%
\begin{figure}
\includegraphics[bb=147 66 453 699,width=1.0\hsize,clip]{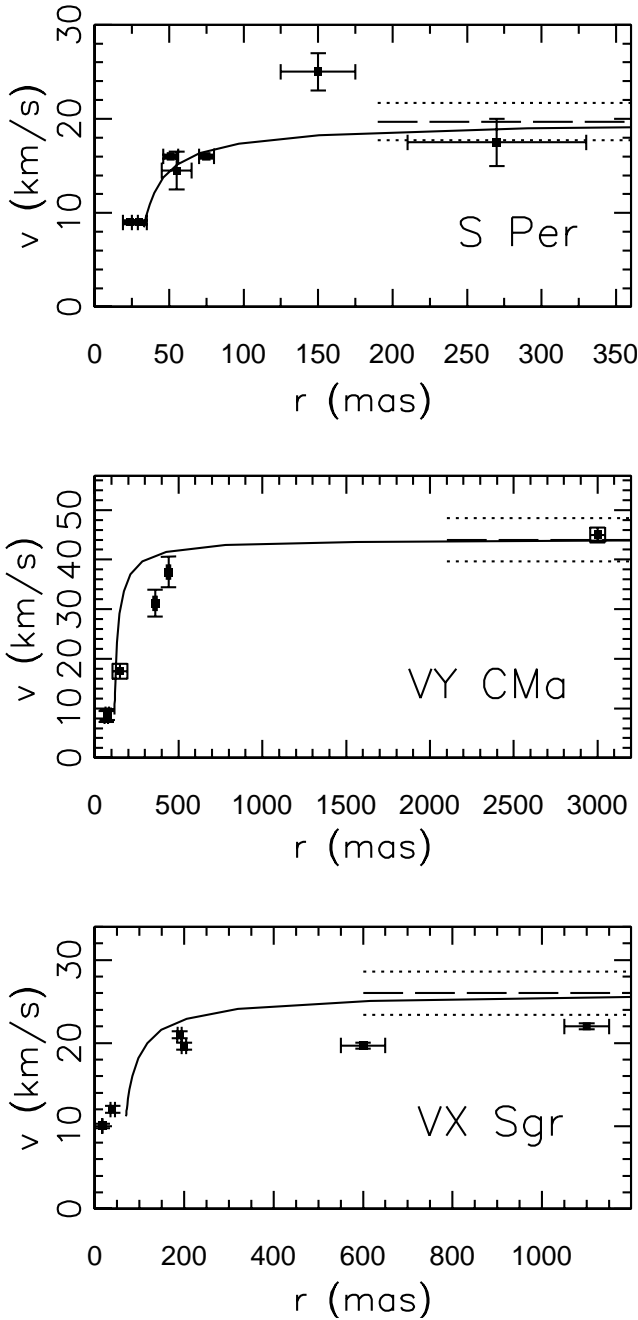}

\caption{Observations and modeling of the velocity profiles of three O-rich
supergiants.  The profiles are traced by various masers observations, shown as
symbols with error bars (from Richards \& Yates 1998).  The model predictions
from eq.\ \ref{eq:velprof} are shown with the solid lines.  The terminal
velocity \vf\ is taken from CO observations, marked by dashed lines with the
1$\sigma$ uncertainty range marked by dotted lines. \label{fig:3starsVelProf}}
\end{figure}
%%%%%%%%%%%%%%%%%%%%%%%%%%%%%%%%%%%%%%%%%%%%%%%%%%%%%%%%%%%%%%%%%%%%%%%%%%%%%

We compare these model predictions with the spatially-resolved maser
observations of three O-rich supergiants reported by Richards \& Yates (1998).
The combined measurements of SiO, H$_2$O and OH masers trace the velocity
profile over three orders of magnitude of radial distance from just outside the
stellar atmosphere. Comparison with the theoretical profile in equation
\ref{eq:velprof} requires three input parameters. The terminal velocity \vf\ is
measured directly in CO observations because most of the CO emission comes from
the outer parts of the envelope, far beyond the acceleration zone. Next, we
constrain the optical depth, $\tV$, which controls the power-law index $k$, by
fitting the spectral energy distributions, as described below in
\S\ref{sec:SED} (its effect on $v(\theta)$ is minor). Finally, we determine the
angular scale $\theta_0$ from eq.\ \ref{eq:theta0} and the measured \Fbol. In
using this relation we assumed \Tc\ = 700 K for all three stars, but the entire
range 600--800\,K is consistent with the data. All relevant parameters for the
three stars are listed in table~\ref{tab:stars} and the resulting $v(\theta)$
are shown in figure \ref{fig:3starsVelProf}. With parameters constrained only
by photometric and CO observations, the model successfully predicts the
velocity profiles measured from the independent maser observations. The good
agreement is obtained even though the model profiles were not fitted to the
data with any adjustable free parameters. This supports the prediction given by
eq.~\ref{eq:velprof}, as well as the $\theta_0$ vs. \Fbol\ relationship given
by eq.~\ref{eq:theta0}.

%%%%%%%%%%%%%%%%%%%%%%%%%%%%% Table 2 %%%%%%%%%%%%%%%%%%%%%%%%%%%%%
\begin{table}
\begin{center}
\begin{tabular}{lrrrc}
\hline
 Star:              &   S Per      & VY CMa       &   VX Sgr    & source  \\
\hline \hline
{\bf Measured:}           &              &              &             &         \\
 Spec. Type         &  M3I         & M3/M4II      & M5/M6III    &     a   \\
  $D$ (kpc)         &   2.3        &   1.5        &   1.5       &     b   \\
     $M$ (\Mo)      &    20        &    50        &    10       &     b   \\
   \Mdot (\Myr)     &   2.7E$-$5   &  1.0E$-$4    &  2.5E$-$5   &     b   \\
  \Fbol\ (W/m$^2$)  & 5.1E$-$10    & 6.6E$-$9     & 2.3E$-$9    &     c   \\
    $L$ (\E4\ \Lo)  &    8.1       &    44        &    16       &     d   \\
$v^{\rm CO}$ (\kms) &    19.7      &    44.0      &     26.0    &     e   \\
%  IRAS name        & 02192+5821   & 07209$-$2540 &18050$-$2213 &    \\
  LRS class         &   26         &    24        &    26       &     f \\
  $[25]-[12]$       &  $-$0.16     &  $-$0.17     &  $-$0.29    &     f \\
  $[60]-[25]$       &  $-$0.76     &  $-$0.66     &  $-$0.72    &     f \\
\hline
{\bf Model:}              &              &              &             &    \\
$\theta_0$ (mas)    &    30        &    110       &      65     &    \\
      \tV           &     3.2      &    18.0      &      5.6    &    \\
\hline
\end{tabular}

\caption{Measured and model-derived parameters for three stars with spatially
resolved kinematic observations from Richards \& Yates (1998). Sources are: a)
HIPPARCOS database; b) Richards \& Yates (1998), and references therein; c)
from integrating observations listed in CIO (Gezari et al. 1993); d) $L$ is
calculated from $4\pi D^2\Fbol$; e) M. Rupen (private communication); f) IRAS
database. \label{tab:stars}
}
%\smallskip
\end{center}
\end{table}
%%%%%%%%%%%%%%%%%%%%%%%%%%%%%%%%%%%%%%%%%%%%%%%%%%%%%%%%%%%%%%%%%%%

\section{Correlations among Dynamical Quantities}

Specifying the dust chemical composition determines its optical properties,
with the relevant parameters listed in table \ref{tab:dust}. An additional,
independent property is the dust abundance, conveniently parametrized with the
dust cross-sectional area per gas particle upon grain condensation
\eq{\label{eq:s22}
  \pi a^2 {\nd\over\nH}\Big|\sub c = \E{-22}\ss\ \hbox{cm}^{2};
}
that is, \nH\ and \nd\ are the number densities of hydrogen nuclei and dust
particles, respectively, before the drift sets in (see paper I, eq.\ 5). The
cross-section \ss\ is a free parameter that controls the relation between
optical depth, mass-loss rate and luminosity as (paper I, eq.\ 55)
\eq{\label{eq:tauV}
    \tV = \alpha\,\sigma_{22}^{2/3}\,{\Mdot_{-6}^{4/3}\over\DS L_4}
}
where \Msix\ = $\Mdot/(\E{-6}\,\Myr)$, $L_4 = L/(\E4\,\Lo)$ and $\alpha =
0.34\QV(T^4\sub{c3}Q_{\ast}^{-2}\Psi_0^{-1})^{1/3},$ with $T\sub{c3} =
T\sub{c}/(1000\,\rm K)$. While \ss\ has no bearing upon the shape of the radial
profile $v/\vf$, it sets the scale of the outflow terminal velocity, which is
related to \tV\ and \Mdot\ as (paper I, eq.\ 52)
\eq{\label{eq:universal}
    \vf = A\,\frac{\Mdot_{-6}^{1/3}}{(1 + \tV)^{1/2}}\ \kms,
}
where $A = 67.5(T^4\sub{c3}Q_{\ast}\Psi_0^{-1})^{1/3}\times\sigma_{22}^{2/3}$.
Combining eqs.\ \ref{eq:tauV} and \ref{eq:universal}, it is evident that when
\Mdot\ is varied at a fixed $L$, the terminal velocity reaches maximum when
\tV\ = 1, i.e., at a mass loss rate $\Mdot(v\sub{max}) \propto  L^{3/4}$. From
this maximum, \vf\ decreases when \Mdot\ is either increasing or decreasing
away from $\Mdot(v\sub{max})$. This decrease reflects the role of the drift in
one direction, reddening the other.

\subsection{ Drift-dominated regime }
\label{Sec:Asection}

%%%%%%%%%%%%%%%%%%%% %%% %%%%%%%%%%%%%%%%%%%%%%%%%%%%%%%%%%%%%%%%%%%%%
\begin{figure}
\includegraphics[bb=20 57 592 609,width=1.0\hsize,clip]{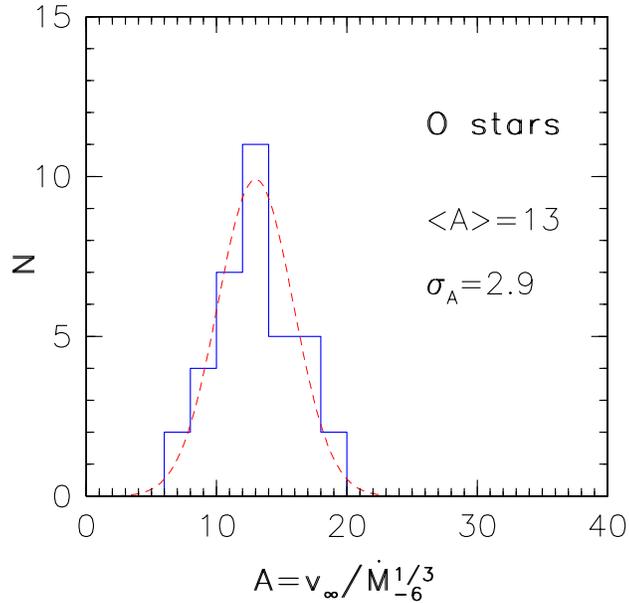}

\caption{Histogram of $A = \vf/\Mdot^{1/3}$ (in \kms) for 36 oxygen-rich stars
with $\tV < 1$ outflows, with data taken from Young (1995). The mean $\langle A
\rangle$ and its dispersion $\sigma_A$ are listed, and the dashed line shows
the Gaussian drawn with these parameters. \label{fig:Aoxygen}}
\end{figure}
%%%%%%%%%%%%%%%%%%%%%%%%%%%%%%%%%%%%%%%%%%%%%%%%%%%%%%%%%%%%%%%%%%%%%%%%%%%%%

%%%%%%%%%%%%%%%%%%%% %%% %%%%%%%%%%%%%%%%%%%%%%%%%%%%%%%%%%%%%%%%%%%%%
\begin{figure}
\includegraphics[bb=20 57 592 609,width=1.0\hsize,clip]{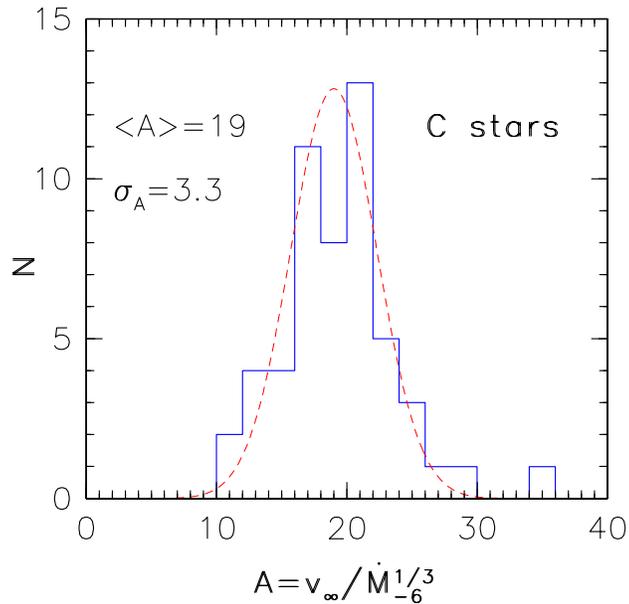}

\caption{Same as figure \ref{fig:Aoxygen}, except for 60 carbon-rich stars. The
data are taken from Olofsson et al. (1993). \label{fig:Acarbon}}
\end{figure}
%%%%%%%%%%%%%%%%%%%%%%%%%%%%%%%%%%%%%%%%%%%%%%%%%%%%%%%%%%%%%%%%%%%%%%%%%%%%%

When \Mdot\ decreases away from $\Mdot(v\sub{max})$ so that \tV\ becomes less
than unity, the dust and gas decouple and the velocity decreases too. In this
drift dominated regime\footnote{For grains smaller than about 0.01 \mic,
the dust drift can be negligible even when $\tV < 1$; for discussion see Section 
C1 in paper I.}, eq.\ \ref{eq:universal} becomes
\eq{
\label{eq:Young}
    \tV < 1: \quad  \vf \simeq A\,\Mdot_{-6}^{1/3}\ \kms.
}
It is remarkable that, even though the wind is driven by radiation pressure,
its velocity is {\it independent of luminosity} for small \Mdot\ (optically
thin limit; see also \S4.2 in paper I). This surprising result was discovered
observationally by Young (1995) in a survey of 36 nearby Mira variables with
low mass-loss rates.  Young found a clear, strong correlation between outflow
velocity and mass-loss rate, but independent of luminosity. His correlation can
be parametrized as $\vf \propto \Mdot^{1/3.35}$, in agreement with eq.\
\ref{eq:Young} if the observational errors for the power-law index are at least
10\%, which is plausible. Figure~\ref{fig:Aoxygen} presents Young's data as a
histogram of $\vf/\Mdot^{1/3}$. The figure also lists the distribution mean
$\langle A \rangle$ and its root-mean-square scatter $\sigma_A$, and plots the
Gaussian with these parameters. Figure~\ref{fig:Acarbon} presents a similar
analysis of C-rich stars with small optical depths, with data from Olofsson et
al. (1993). In the case of Young's data we did not introduce any cuts since the
sample is dominated by optically thin envelopes, as can be seen from figure
\ref{fig:basicO}. On the other hand, the 63 C-rich stars in the Olofsson et al
data include three objects with $\tV > 1$, as is evident in figure
\ref{fig:basicC}, and these were excluded from the histogram. Each histogram
shows a pronounced peak---in agreement with eq.\ \ref{eq:Young} when the dust
properties do not vary much within the sample. The ratio $\sigma_A/\langle A
\rangle$ is $<25\%$ for each sample, a fractional scatter consistent with the
measurement errors ($\sim$10\%  for \vf\ and $\sim$50\% for \Mdot; see Appendix
for details). These strongly peaked distributions affirm the central role of
dust drift at small mass loss rates and indicate a close similarity of dust
properties within each sample.

\subsection{   Dust-to-gas ratios   \label{sec:rdg} }

\begin{figure}
\includegraphics[bb=57 50 510 716, width=1.0\hsize, clip]{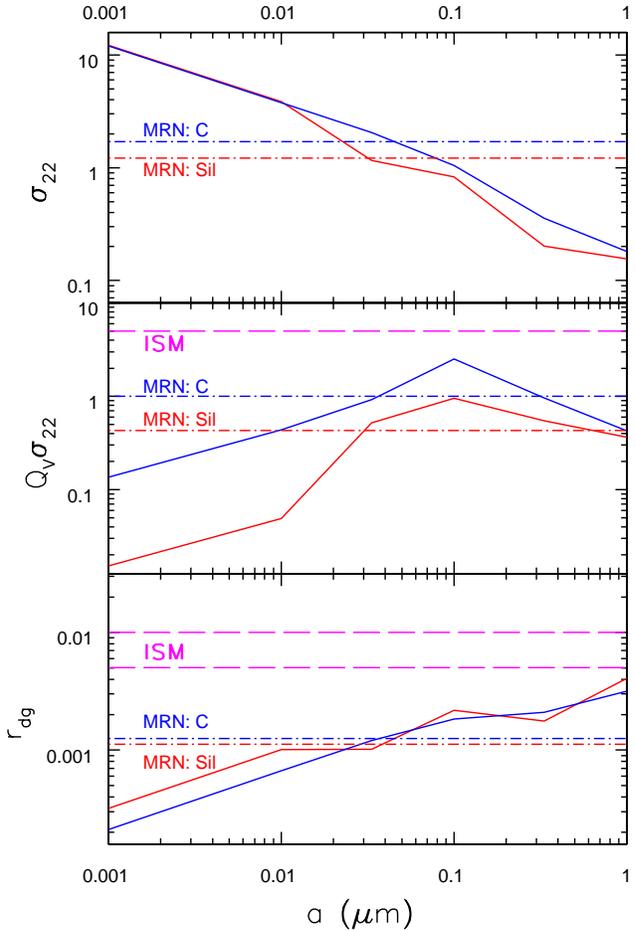}

\caption{Dust abundance in winds around evolved stars and in the Galactic
interstellar medium (ISM). {\em Top panel}: The dust geometric cross-section
per gas particle inferred for dusty winds from eq.~\ref{eq:sig22} as a function
of grain size. Results for amorphous carbon  grains are shown in blue solid
line, for silicate in red. The inferred values for the MRN grain size
distribution (Mathis, Rumpl \& Nordsieck 1977) are shown with horizontal
dot-dashed lines. {\em Mid panel}: Analogous to the top panel, except that it
shows the product $Q_V\ss$, whose value in the interstellar medium is directly
constrained by observations. The ISM value $Q_V\ss = 5$ is marked with the
dashed horizontal line. {\em Bottom panel}: Variation with grain size of the
dust-to-gas mass ratio (eq. \ref{eq:rdg}) in dusty winds. The ISM range is
0.5--1\%, marked by the dashed horizontal lines.
\label{fig:sig22}}
\end{figure}

The velocity scale $A$, 19 \kms\ for C-rich and 13 \kms\ for O-rich stars, is a
fundamental property of dusty winds, derived directly from the data. Adding
assumptions regarding the grain properties enables determination of the dust
geometric cross-section per gas particle from
\eq{
    \ss = \beta\,A^{3/2}
\label{eq:sig22}
}
where $\beta = 1.8\x\E{-3}\Psi_0^{1/2}Q_{\ast}^{-1/2}T^{-2}\sub{c3}$. The top
panel of figure~\ref{fig:sig22} shows the variation with grain size of the
inferred values of \ss\ for amorphous carbon and silicate grains. Remarkably,
the two samples of different type stars produce values of \ss\ that agree to
within 50\% at all grain sizes. In spite of the large differences in
atmospheric and grain properties between O- and C-rich stars, the fraction of
material channelled into dust is such that the dust area per gas particle turns
out to be roughly the same in both.

The quantity closest to \ss\ that is directly determined in observation is the
dust extinction per H column density. The Galactic average yields $N_H/\tV =
1.95\x\E{21}$ cm$^{-2}$ (e.g., Sparke \& Gallagher 2006), which translates to
\QV\ss\ = 5. The middle panel of figure~\ref{fig:sig22} shows the variation
with grain size of the value of \QV\ss\ inferred from the measured values of
$A$ for the C- and O-rich stars. Galactic interstellar dust contains a mixture
of carbon and silicate grains of various sizes, but the dusty wind results are
significantly below the Galactic average for both types of dust and any grain
size; averaging the results with the MRN (Mathis, Rumpl \& Nordsieck 1977) size
distribution yields the values indicated in the figure with dot-dashed lines.

Another indicator of dust abundance is the dust-to-gas mass ratio \rdg; it is
widely used even though its determination invokes assumptions about molecular
abundances that bring additional uncertainty. Estimates of the Galactic average
of \rdg\ range from \about\ 0.005 (Draine 2009) to 0.01 (Barbaro et al 2004).
Recalling that \ss\ characterizes the base of the outflow where the gas and
dust have the same velocity (prior to the dust drift; see eq.\ \ref{eq:s22}),
the outflow dust-to-gas mass ratio is
\eq{\label{eq:rdg}
    \rdg = {\Mdot\sub{d} \over \Mdot}
         = 2.40\x\E{-3}{\rho\sub{s}\over3\,{\rm g\,cm^{-3}}}
                       {a\over 0.1\mic}\,\ss
}
where $\rho\sub{s}$ is the density of the grain material (3.28 g\,cm$^{-3}$ for
silicate dust, 2.2 g\,cm$^{-3}$ for carbon grains).  The bottom panel of
figure~\ref{fig:sig22} shows the variation with grain size of the value of
\rdg\ for dusty winds, providing yet another display of two properties noted
above: (1) In spite of the large differences in their formation properties, the
mass fraction of carbon and silicate grains is roughly the same in both types
of stars, and (2) the wind dust abundance is significantly below the Galactic
average.

These results have significant consequences regarding the origin of
interstellar dust. Using either \ss\ or \rdg\ as indicators, the dust abundance
in winds around evolved stars is substantially lower than the Galactic averages
irrespective of grain size or chemical composition. The MRN average for \QV\ss\
in dusty winds is more than a factor of 10 below the Galactic average for
silicate dust, a factor of 5 for carbon. Since the winds fail to produce the
observed ISM values for both types of grains at all sizes, mixture averaging
cannot bring the results to the Galactic values. The implication is that dusty
winds {\em cannot} be the source of all Galactic dust: If all dust were formed
by cycling interstellar gas through stars, with existing grains destroyed
during star formation and reformed around evolved stars, than the Galactic
average could not exceed the dusty winds value. Draine (2009) discusses
additional grain destruction mechanisms in estimating the Galactic dust budget,
and concludes that they could not be offset by formation in stars 
(see also Zhukovska, Gail \& Trieloff 2008). Our results
show that even without considering any processes other than cycling through
stars, {\em most interstellar dust is not stardust and must have formed in the
ISM}, in strong support of Draine's conclusion.

It should be noted that the analysis here provides a robust derivation of the
dust abundance. The parameter $A$ is determined directly from the data, and its
conversion to \ss\ and \rdg\ involves only minimal assumptions about the grain
properties; significantly, no assumptions are made about any molecular
abundances.  The presented results employed \Tc\ = 700 K, and the inferred dust
abundance scales as $T\sub{c}^{-2}$. Varying \Tc\ in its likely range, the dust
abundance would increase at every $a$ by 36\% for \Tc\ = 600 K and decrease by
23\% if \Tc\ = 800 K; such variations have little effect on our conclusions. In
addition, there is no need to consider global balance of dust forming and
destruction processes in the ISM, which are very uncertain. Our conclusion is
derived from consideration of individual stars and therefore it is independent
of, and supports, Draine's arguments.

%%%%%%%%%%%%%%%%%%%% %%% %%%%%%%%%%%%%%%%%%%%%%%%%%%%%%%%%%%%%%%%%%%%%
%%%%%%%%%%%%%%%%%%%%%%%%%%%%%%%%%%%%%%%%%%%%%%%%%%%%%%%%%%%%%%%%%%%%%%%%%%%%%

\subsection{ Optically thick winds }

\begin{figure}
\includegraphics[bb=80 40 532 609,width=1.0\hsize,clip]{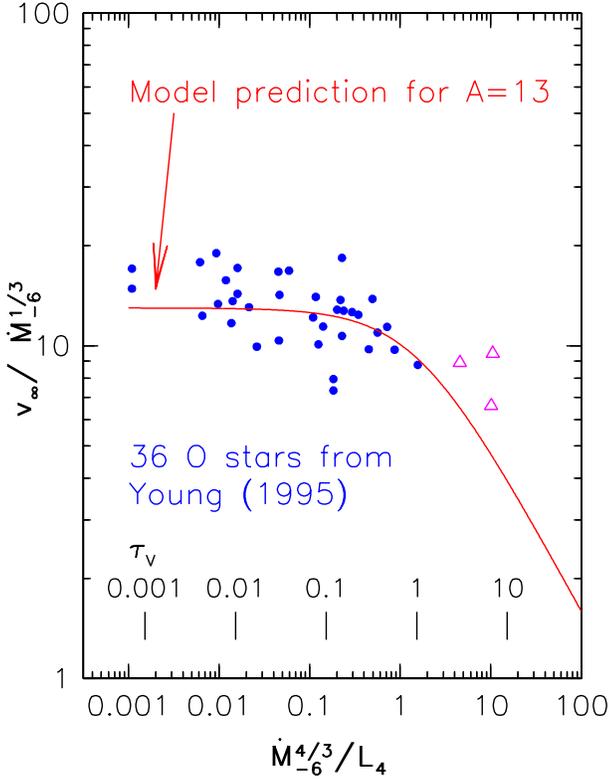}

\caption{A comparison of the model prediction for \vf\ (in \kms) with
observations of O-stars. The line plots the relationship in eq.\
\ref{eq:universal}, with the $A$ value determined from the small--\Mdot\ limit
(figure \ref{fig:Aoxygen}). The data are from Young (1995, dots) and Richards
\& Yates (1998, triangles). \label{fig:basicO}}
\end{figure}

\begin{figure}
\includegraphics[bb=80 40 532 609,width=1.0\hsize,clip]{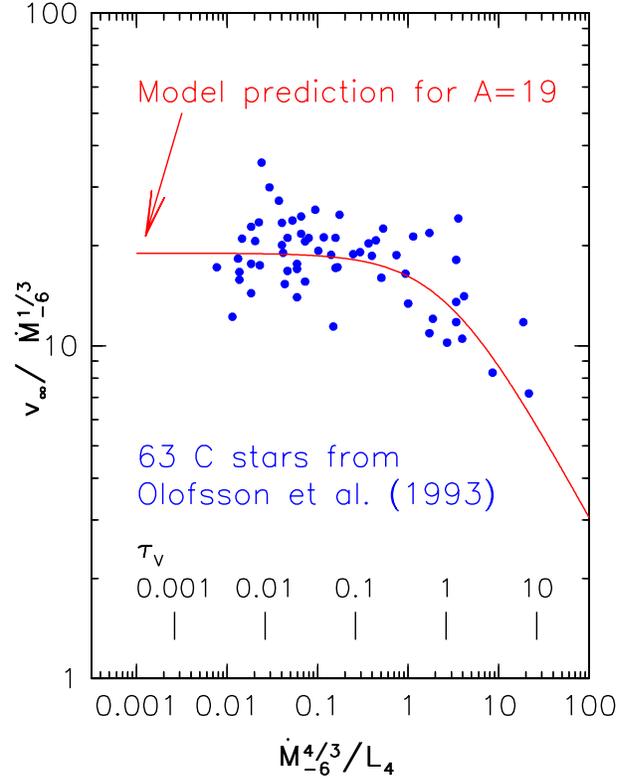}

\caption{Same as figure \ref{fig:basicO}, except for C-stars. The $A$ value is
from figure \ref{fig:Acarbon}, the data are from Olofsson et al.\ (1993).
\label{fig:basicC}}
\end{figure}

As \Mdot\ increases away from $\Mdot(v\sub{max})$, the wind becomes optically
thick and reddening degrades the efficiency of the radiation pressure force,
thus the velocity again decreases. The wind velocity is no longer independent
of luminosity. Instead, from equations \ref{eq:universal} and \ref{eq:tauV} it
follows that when $\tV > 1$, the final velocity is proportional to
$L^{1/2}/\Mdot^{1/3}$. Therefore, the ratio $\vf/\Mdot^{1/3}$ is no longer
constant, instead it decreases as $\tau_{\rm V}^{-1/2} \propto
(\Mdot^{4/3}/L)^{-1/2}$ from its value in the optically thin regime.
Figures~\ref{fig:basicO} and \ref{fig:basicC} show the comparison of model
predictions with observations, by plotting the data and the relationship in
eq.\ \ref{eq:universal} in the $\vf/\Mdot^{1/3} - \Mdot^{4/3}\!/L$ plane. Since
the single free parameter $A$ is taken from the histograms in figures
\ref{fig:Aoxygen} and \ref{fig:Acarbon}, {\em the agreement between model and
data displayed in figures~\ref{fig:basicO} and \ref{fig:basicC} is obtained
without adjusting any parameters}; the coefficient $A$ controls both the
constant asymptotic value of $\vf/\Mdot^{1/3}$ in the small $\Mdot^{4/3}/L$
regime as well as the value of $\Mdot^{4/3}/L$ at which $\vf/\Mdot^{1/3}$
begins to decrease, thus this agreement represents another test of the model.
This test would be much stronger if the samples contained more stars with very
large values of \tV\ (i.e., $\Mdot^{4/3}/L$) so that the transitions from
optically thin to thick regimes were better defined. We note that the samples
are dominated by optically thin envelopes, thus our above estimates of the mean
$A$ are not appreciably biased by the few stars with moderate optical depths.

%%%%%%%%%%%%%%%%%%%% %%% %%%%%%%%%%%%%%%%%%%%%%%%%%%%%%%%%%%%%%%%%%%%%
%%%%%%%%%%%%%%%%%%%%%%%%%%%%%%%%%%%%%%%%%%%%%%%%%%%%%%%%%%%%%%%%%%%%%%%%%%%%%

%%%%%%%%%%%%%%%%%%%% %%% %%%%%%%%%%%%%%%%%%%%%%%%%%%%%%%%%%%%%%%%%%%%%
%%%%%%%%%%%%%%%%%%%%%%%%%%%%%%%%%%%%%%%%%%%%%%%%%%%%%%%%%%%%%%%%%%%%%%%%%%%%%

\section{Physical Domain of Dusty Winds}
\label{sec:domain}

The relation $\Mdot\vf \le L/c$ has often been invoked as the momentum
conservation bound on radiatively driven mass loss rates, even though the
mistake in this application when $\tV > 1$ has been pointed out repeatedly
(e.g.\ Ivezi\'c \& Elitzur 1995). Instead, the proper form of momentum
conservation is $\Mdot\vf = \tF L/c$, where \tF\ is the flux-averaged optical
depth. Since \tF\ can exceed unity for plausible values of \tV\ (see paper I,
eq.\ 57 and figure 6), momentum conservation does not impose a meaningful
constraint on dusty winds. Instead, the constraints come from force
considerations---the radiative outward force must exceed everywhere the
gravitational pull of the star with mass $M$. This condition breaks down at the
two ends of the mass-loss-rate range, where the outward force is reduced for
two different reasons: at very low \Mdot\ the gas--dust momentum coupling
weakens thus reducing the force on the gas, and at very high \Mdot\ the
coupling to the radiative force diminishes because of the enhanced reddening
that follows increased obscuration. In paper I (see sec.\ 5, in particular
figure 7) we derive the resulting phase-space boundaries with the aid of the
appropriate scaling variables. Here we reproduce the results in terms of the
system physical parameters.

\begin{figure}
\includegraphics[width=1.0\hsize,clip]{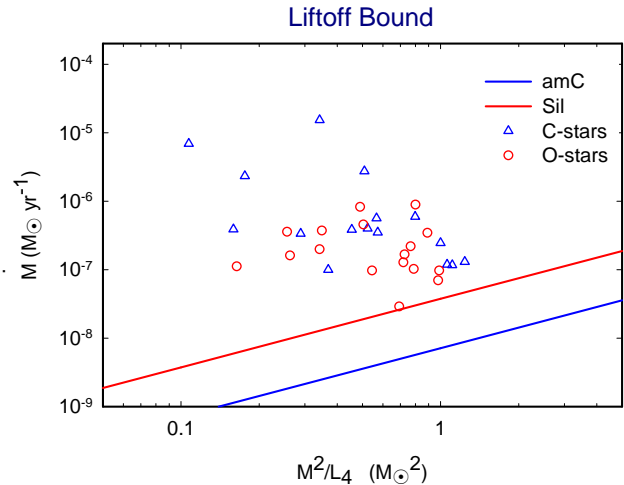}

\caption{Winds driven by radiation pressure on dust should fall above the
liftoff bounds shown by lines. The observations for stars with mass
estimates available from SIMBAD are shown by symbols. These estimates
may have large uncertainties due to heterogeneous sources.
\label{fig:thinBound}}
\end{figure}

Figure~\ref{fig:thinBound} shows the lower bound on \Mdot, arising from the
requirement that radiation pressure on the dust should provide sufficient force
on the gas to generate a net acceleration at the base of the outflow. This
liftoff condition sets a lower bound on \Mdot, proportional to $M^2/L$ (paper
I, eq.\ 69); below this minimal $\Mdot_{\rm min}$ the grains are ejected
without dragging the gas with them because the density is too low for efficient
gas--dust coupling. As a condition on the wind initiation, this bound is the
most uncertain part of our solution. The result $\Mdot_{\rm min} \propto M^2/L$
is reasonably secure (a similar relation was noted by Habing et al 1994), but
the proportionality constant can be determined accurately only from a more
complete formulation that handles properly grain growth and the wind launching
mechanism.

\begin{figure}
\includegraphics[width=1.0\hsize,clip]{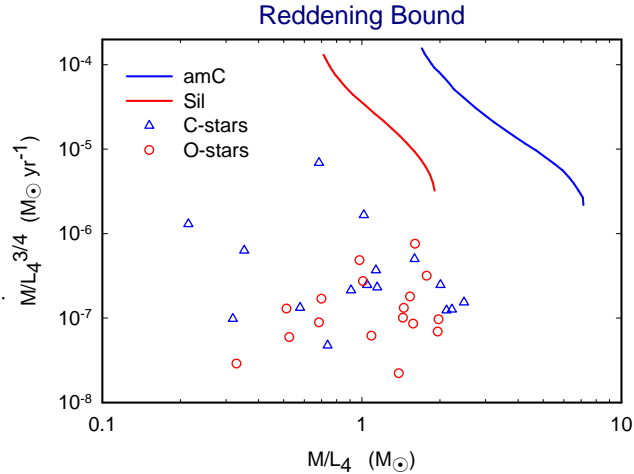}

\caption{Winds driven by radiation pressure on dust should fall below the
reddening bounds shown by lines. The observations are shown by symbols.
\label{fig:thickBound}
}
\end{figure}

The bound shown in figure \ref{fig:thickBound} reflects the weakening of
radiative coupling in optically thick winds because of the radiation reddening.
For any Eddington ratio (the horizontal axis), the figure shows the upper limit
on the wind optical depth \tV\ ($\propto \Mdot^{4/3}/L$; see eq.\
\ref{eq:tauV}), determined from the full numerical solution of the self-similar
problem; the vertical axis units were chosen to vary linearly with \Mdot.
Although some stars have $\Mdot\vf \ga L/c$, none violate the upper bound set
by proper solution of the dusty wind problem.

%%%%%%%%%%%%%%%%%%% %%% %%%%%%%%%%%%%%%%%%%%%%%%%%%%%%%%%%%%%%%%%%%%%
%%%%%%%%%%%%%%%%%%%%%%%%%%%%%%%%%%%%%%%%%%%%%%%%%%%%%%%%%%%%%%%%%%%%%%%%%%%%

%%%%%%%%%%%%%%%%%%% %%% %%%%%%%%%%%%%%%%%%%%%%%%%%%%%%%%%%%%%%%%%%%%%
%%%%%%%%%%%%%%%%%%%%%%%%%%%%%%%%%%%%%%%%%%%%%%%%%%%%%%%%%%%%%%%%%%%%%%%%%%%%

\section{The Wind IR Emission}
\label{sec:SED}

Given the grain properties, the dusty wind problem requires three independent
input parameters, which can be selected as the initial velocity, the Eddington
ratio and the overall optical depth (paper I). However, the dependence on the
first two is limited to the immediate vicinity of the boundary of the
phase-space for physical solutions; away from that boundary, dusty winds are
described by a set of similarity functions of the single independent variable
\tV. Figures \ref{fig:thinBound} and \ref{fig:thickBound} show that most stars
are located well inside the allowed region of phase space and thus should be
well described by \tV\ alone. Therefore, we should be able to characterize
every property of the wind IR emission (SED, colors, etc.) with \tV\ only.

\begin{figure}
\includegraphics[bb=87 60 510 704,width=1.0\hsize,clip]{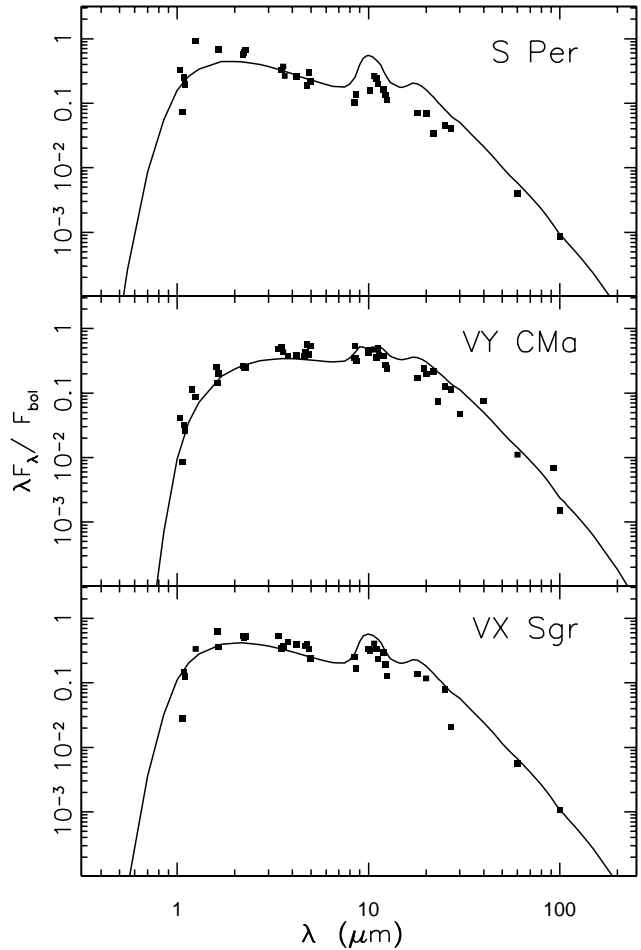}

\caption{The observed SEDs (symbols) and model fits (lines) for the three
O-rich supergiants whose velocity profiles are shown in figure
\ref{fig:3starsVelProf}. Data are taken from Gezari et al. (1993).
The model parameters are listed in Table 2.
\label{fig:3starsSED}}
\end{figure}

We employed the code DUSTY (Ivezi\' c, Nenkova \& Elitzur 1999) to compute the
SED of each of the three oxygen-rich stars whose velocity profiles, as well as
the approximate analytic solution of eq.\ \ref{eq:velprof}, are shown in figure
\ref{fig:3starsVelProf}. The DUSTY calculations were done using the code's
option ``analytic radiatively driven wind'', which computes the density profile
from the same approximate analytic solution (for details, see the DUSTY
manual). Figure \ref{fig:3starsSED} shows the model SEDs, fitted to the data
with the single free parameter \tV. The same models that successfully explain
the velocity profiles measured from maser observations also produce
satisfactory fits to SED of each star. The best-fit values for \tV\ are listed
in table~\ref{tab:stars}. Another example where SED fits provided successful
predictions for the spatially resolved velocity profile is W Hya. Zubko \&
Elitzur (2000) fitted simultaneously the SED and the velocity profile deduced
from observations of the CO thermal emission and various masers. Significantly,
neither $L$ nor \Mdot\ or dust-to-gas ratio were input parameters in that
model. Instead, these quantities were derived from general self-similarity
relations after the SED fitting results were supplemented by the distance and
velocity scales.

Successful SED fitting is not limited to stars with silicate dust. We have
shown elsewhere that the SED of the very dusty carbon-rich star IRC+10216 can
be fitted as a function of pulsation phase by simply varying optical depth
(Ivezi\'{c} \& Elitzur 1996).

%%%%%%%%%%%%%%%%%%%% %%% %%%%%%%%%%%%%%%%%%%%%%%%%%%%%%%%%%%%%%%%%%%%%
%%%%%%%%%%%%%%%%%%%%%%%%%%%%%%%%%%%%%%%%%%%%%%%%%%%%%%%%%%%%%%%%%%%%%%%%%%%%%

%%%%%%%%%%%%%%%%%%%% %%% %%%%%%%%%%%%%%%%%%%%%%%%%%%%%%%%%%%%%%%%%%%%%
\begin{figure}
\includegraphics[bb=60 60 518 715,width=1.0\hsize,clip]{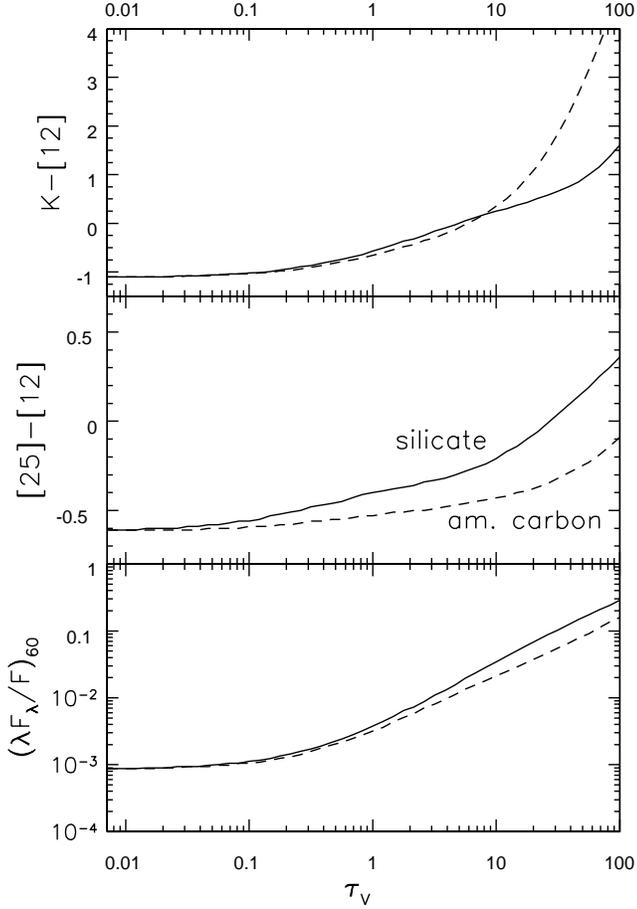}

\caption{The dependence of various IR colors on optical depth $\tau_V$, for two
types of dust. Predictions for other bandpasses, or for different dust
properties, can be easily generated with the DUSTY code. \label{fig:IR_tauV}}
\end{figure}
%%%%%%%%%%%%%%%%%%%%%%%%%%%%%%%%%%%%%%%%%%%%%%%%%%%%%%%%%%%%%%%%%%%%%%%%%%%%%

\subsection{A Comment on Estimating Mass-loss Rate from IR Observations}

The infrared emission from dust is related to the gas mass-loss rate and may
offer observationally convenient way to estimate the latter. The literature is
abundant with various proposed expressions that relate IR  observables and
mass-loss rate  (e.g. van Loon 2008, and references therein; see also van der
Veen \& Rugers 1989). However, in addition to heterogeneous data sets and
methods (see Appendix for a summary), it is often unclear what assumptions are
made, and what variables are considered to be independent. A key point of our
analysis is that there are only two {\it independent} relations between various
relevant quantities, and they reflect energy and momentum conservation
(eqs.~\ref{eq:tauV} and \ref{eq:universal}). Given these two relationships,
only two quantities can be derived, and all others have to be assumed, or
measured. Thanks to its scaling properties, the problem of connecting gas
mass-loss rate to IR observables can be decoupled into two independent steps:
determination of \tV\ from IR observables, and relating \tV\ to dynamical
quantities, including gas mass-loss rate.

\begin{figure}
\includegraphics[bb=35 60 537 724,width=1.0\hsize,clip]{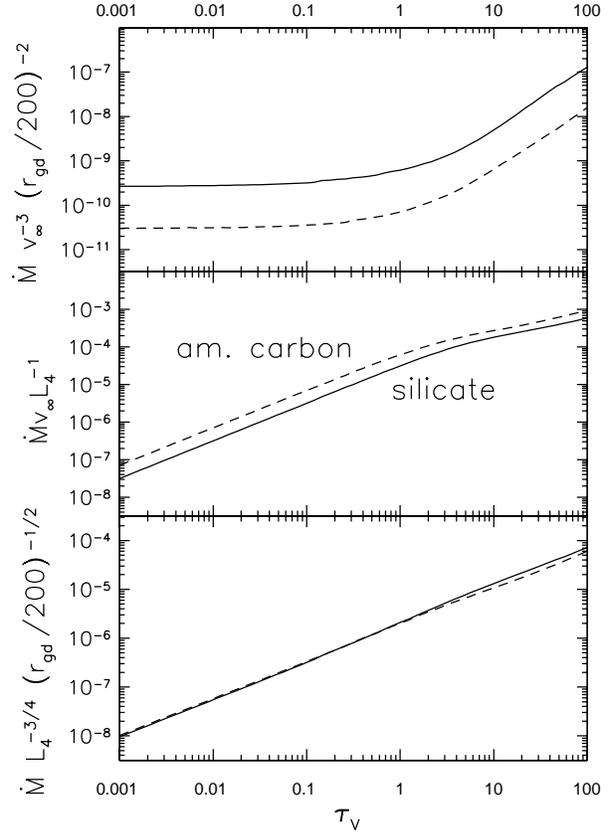}

\caption{Model predictions for correlations between various dynamical
quantities and optical depth \tV\ (only two relations are independent). {\em
Top panel}: Plot of the relation in eq.~\ref{eq:universal}. {\em Mid panel}:
Ratio of mechanical to radiative momentum. {\em Bottom panel}: Plot of the
relation in eq.~\ref{eq:tauV}. In all panels \vf\ is in \kms, \Mdot\ in
\Mo\,yr$^{-1}$. The gas-to-dust mass ratio is defined as $r_{gd} = 1/r_{dg}$
(see eq.~\ref{eq:rdg}). \label{fig:dyn_tauV}}
\end{figure}

The most robust and accurate method for estimating \tV\ from observations is
fitting of a well-sampled SED. When the data are sparse, various IR colors can
also be used, albeit with deteriorating accuracy. Examples of such
relationships are show in figure~\ref{fig:IR_tauV}. Given an estimate of \tV,
the model uniquely predicts correlations between \tV\ and various combinations
of dynamical quantities formed using \vf, \Mdot, $L$, and \ss. Examples of such
relations are shown in figure~\ref{fig:dyn_tauV} (based on eqs.~\ref{eq:tauV}
and \ref{eq:universal}, as well as $\Mdot \vf = \tau_F L/c$, with the latter
derived from, and not independent of, the first two expressions). The most
appropriate expression to use for estimating gas mass-loss rate depends on the
available data. For example, in case of the LMC studies, where outflow velocity
is available for a much smaller number of stars than photometry,
 eq.~\ref{eq:tauV} provides a superior approach because it involves only SED fitting
(distance to the LMC can be considered well constrained in this context): for fixed
values of $\alpha$ and $\sigma_{22}$, only \tV\ and $L_4$ are required to estimate
\Mdot. The values for $\alpha$ and $\sigma_{22}$ (see definitions after 
eq.~\ref{eq:tauV}) can be taken as equal to the Galactic
case (Table 1), or could be determined as in Section~\ref{Sec:Asection} using a
sample of LMC stars with outflow velocity and mass-loss rate measurements.

%%%%%%%%%%%%%%%%%%%% %%% %%%%%%%%%%%%%%%%%%%%%%%%%%%%%%%%%%%%%%%%%%%%%
%%%%%%%%%%%%%%%%%%%%%%%%%%%%%%%%%%%%%%%%%%%%%%%%%%%%%%%%%%%%%%%%%%%%%%%%%%%%%

\section{DISCUSSION}

We have presented here a summary of the Elitzur \& Ivezi\'{c} (2001) similarity
solution in terms more suitable for direct comparison with observations. The
correlation of \vf\ and \Mdot\ in optically thin winds (eq.\ \ref{eq:Young}),
first discovered observationally by Young (1995), emerges as a fundamental
property of radiatively driven dusty winds and a powerful tool in the analysis
of their data. Our own analysis in \S\ref{Sec:Asection} uncovers two major new
results: (1) The peaked distributions in figures \ref{fig:Aoxygen} and
\ref{fig:Acarbon} indicate there is little variation in dust properties among
O-rich and C-rich stars, and (2) the dust cross section per gas particle (\ss)
and the dust-to-gas mass ration (\rdg) are essentially the same for the two
classes. It is hard to think of another method that could produce these
conclusions with a similar level of confidence. Both of these results present a
major challenge for theoretical studies of dust formation.

The relatively small scatter around Young's correlation suggests also a new
method for determining mass-loss rates in optically thin outflows from the
relation
\eq{
    \Mdot = \E{-6}\left(\vf\over A\right)^3 \ \Myr,
}
where \vf\ is in \kms\ and $A$ is from \S\ref{Sec:Asection}. Here \Mdot\ is
determined from the single measurement of \vf, without any assumptions about
dust abundance. In addition, there is no distance dependence, nor any need for
complex modeling. The scatter in this relation is expected to be \about\
50-70\%, no worse than any other method for determining mass loss rates (see
Appendix A). The only restriction is that the wind optical depth at visual must
be less than unity, a condition that is easy to verify observationally. Of
course, the absolute uncertainty of \Mdot\ scale is inherited from Young's
(1995) calibration of his CO measurements, via the values of the $A$ parameter.

In the broader context of Galactic ISM dust, our analysis represents
independent support for Draine's (2009) conclusion that most ISM dust particles
were formed in situ, rather than produced in AGB winds. If verified, this
conclusion would have important consequences for our understanding of galaxy
evolution.

Our model captures all the scaling relationships among the main observable
quantities that follow from energy and momentum conservation. These
correlations can be used to estimate quantities whose measurements might not be
available, notably the mass-loss rate, and for modeling the impact of AGB stars
on their host galaxies. An example is the recent extensive modeling study by
Marigo et al. (2008). Our results are also suitable for analysis of massive
data sets such as the recent SAGE survey of the LMC (Meixner et al. 2006). We
emphasize that our derived values of \ss\ and \rdg\ are directly proportional
to the $\vf/\Mdot^{1/3}$ ratio in the optically thin domain. If one wished to
compare \rdg\ for two populations of stars, say from the Galaxy and the LMC, a
robust method is to compare the $\vf/\Mdot^{1/3}$ distributions for samples
verified to be optically thin using spectral energy distribution.

Despite the successful confrontation of our model with observations, reality is
more complex. The model assumes steady state and spherical geometry, but there
is evidence that in some stars mass-loss rate is not steady (see, e.g.,
Marengo, Ivezi\'c \& Knapp 2001, and references therein) and that some contain
an additional bipolar component (see, e.g., Vinkovi\'c et al 2004, and
references therein). In addition, the physical and chemical details at the base
of the acceleration region are not included in our model. It would be prudent
to compare predictions for the velocity spatial profile and relationships among
dynamical and spectral quantities produced by more elaborate models, e.g.,
Fleischer, Winters \& Sedlmayr 1999, H\"{o}fner 1999 and Dorfi et al 2001. Such
a comparison would help identify which features in these complex models are
simply direct consequences of energy and momentum conservation, and which are
unique to detailed modeling of various physical and chemical effects.

\section*{Acknowledgments}

We thank Jill Knapp and Bruce Balick for illuminating discussions. The partial
support of NASA and NSF is gratefully acknowledged.

\appendix

\section{        Summary of Observations   }

Here we summarize the main measuring methods for most relevant
observables, with emphasis on their accuracy and scaling with
distance.

\subsection{      Dynamical Quantities        }

Observational methods for determining mass-loss rate and outflow velocity were
analyzed and compared by van der Veen \& Rugers (1989, hereafter vdVR). More
detailed discussions are presented by Habing (1996), Olofsson (1996, 1997) and
Wallerstein \& Knapp (1998). There are three widely employed methods:
\begin{itemize}
\item
The strength and shape of thermal CO line profiles contain information about
the outflow velocity and the total amount of CO in the circumstellar shell. The
shape and width of the line profile constrains the outflow velocity in an
almost model-independent way. With the current observational capabilities and
moderate signal-to-noise ratios, the outflow velocity can be constrained to
better than 10\%. Most of the CO emission comes from the outer parts of the
envelope, and thus the measured velocity usually corresponds to the final
outflow velocity.

The relationship between the implied CO mass and the directly observed
quantities is a complex model-dependent function. The expressions most often
used in data analysis were derived by Knapp \& Morris (1985). The
transformation from the CO mass to gas mass-loss rate requires further
assumptions about the outflow, its geometry, and the CO-to-gas ratio. Assuming
a steady-state outflow and spherical geometry, vdVR derive expressions that can
be used to estimate gas mass-loss rate with an uncertainty of about a factor
2-5. More accurate estimates can be obtained by detailed modeling of individual
sources, but probably not significantly better than a factor of 2. We note that
the mass-loss rate estimate scales with $D^2$, where $D$ is the source
distance.

\item

The OH(1612MHz) maser line profile contains information about the outflow
velocity and the amount of OH molecules. The line profile usually has two
well-defined peaks whose velocity separation typically constrains the outflow
velocity to better than 10\%. This estimate is less model-dependent than the
estimate based on the CO line profile, but unfortunately can only be used for
oxygen-rich (O) stars.

Baud \& Habing (1983) proposed a simple model-dependent relation that can be
used to estimate the mass-loss rate from the peak flux of OH emission. As
pointed out by vdVR, the uncertainty of this estimate can be as large as a
factor of 5 due to strong temporal variations in the OH emission strength.
Intrinsic accuracy of this method is probably not much better than a factor of
2 due to a number of assumptions made, and due to uncertain values of the
OH-to-gas ratio. We note that the mass-loss rate estimate proposed by Baud \&
Habing scales with $D$.

\item

The third widely employed method for estimating mass-loss rate is based on
infrared observations of dust emission. The proposed expressions (Herman {\em
et al.} 1986, Jura 1987) relate the observed emission to the optical depth,
which in turn is assumed to encode information about the gas mass-loss rate and
outflow velocity. Similarly to the above two methods that rely on the
assumptions about the CO-to-gas and OH-to-gas ratios, the infrared-based
mass-loss rate estimate is greatly affected by uncertain dust-to-gas ratio.
Depending on numerous additional assumptions employed by various authors, the
distance dependence of the infrared mass-loss rate estimates varies from
proportional to $D^2$, to no dependence at all.

Various observables that relate infrared emission to optical depth have been
also been widely utilized. Both Herman {\em et al.} and Jura assume that the 60 \mic\ dust
emission is optically thin, leading to $\tau \propto F_{60}/F_{bol}$, where
$F_{60}$ is the IRAS flux at 60 \mic\ and $F_{bol}$ is the bolometric flux. Van
der Veen and Rugers (vdVR) relate the optical depth to the $F_{25}/F_{12}$ flux
ratio, where $F_{12}$ and $F_{25}$ are the IRAS fluxes at 12 and 25 \mic.

One of the most widely used prescriptions for determining mass-loss rate
from IR observations was proposed by Jura (1987). He assumed that the dust
mass-loss rate is proportional to dust outflow velocity and dust optical
depth, and that dust outflow velocity is proportional to gas outflow
velocity. Dust optical depth is assumed to be linearly proportional to the
far-IR flux (IRAS 60 \mic\ bandpass) because both the dust optical depth
and the stellar contribution to the overall flux are very small at such
long wavelengths. Given the importance of dust drift in optically thin
regime discussed in \S\ref{Sec:Asection}, the success of Jura's formula is
very surprising: for small optical depths the ratio of dust and gas
velocities is much larger than unity, which should lead to significant
underestimate of mass-loss rate.

It turns out that there are two effects that offset each other, and the
Jura's expression for \Mdot\ coincidentally produces correct values over a large
dynamic range of mass-loss rate. In addition to neglected dust drift, the
assumption about far-IR flux being dominated by dust emission also breaks
down in optically thin regime. These effects are quantitatively illustrated
in figure~\ref{fig:Jura}: at small mass-loss rates the increase of the
ratio of dust and gas velocities is well matched by the decrease of the
dust emission contribution to the 60 \mic\ flux. As a result, Jura's
mass-loss rate estimate never deviates by more than a factor of 2 from the
median mass-loss rate ratio shown by the dotted line in
figure~\ref{fig:Jura} (this median ratio is $\sim$2.5 and represents an
overall systematic offset of two mass-loss rate scales). Hence, the success of Jura's
formula, which does not incorporate dust drift, is {\it not} an argument
against our model which includes dust drift.

%%%%%%%%%%%%%%%%%%%% %%% %%%%%%%%%%%%%%%%%%%%%%%%%%%%%%%%%%%%%%%%%%%%%
\begin{figure}
\includegraphics[bb=90 188 516 609,width=1.0\hsize,clip]{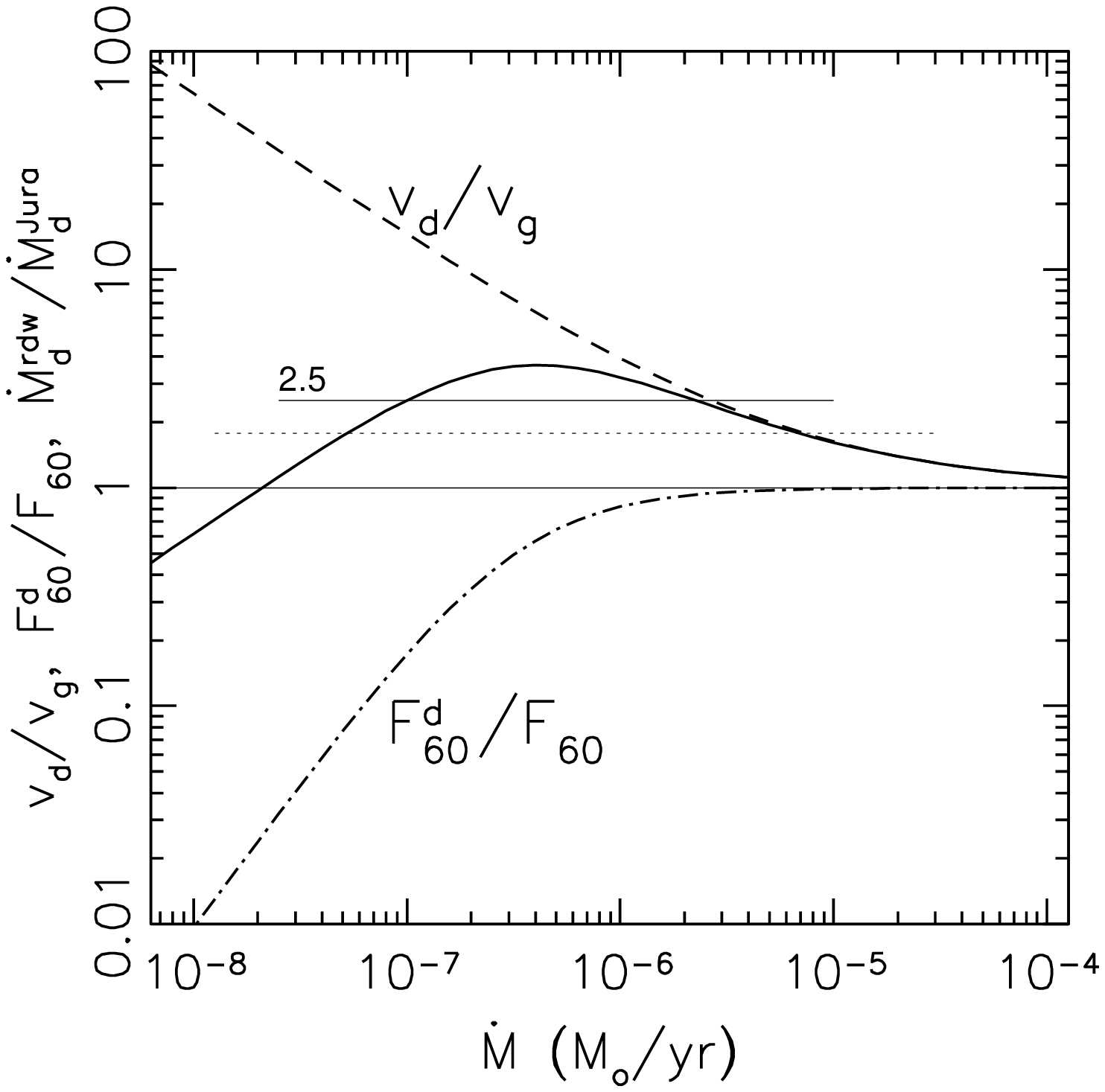}

\caption{The ratio of mass-loss rate given by our model and that derived
using Jura's formula is shown by the solid line, as a function of the
former. The ratio of dust and gas velocities is shown by the dashed line,
and the ratio of 60 \mic\ dust emission and total flux (i.e. including the
stellar contribution) is shown by the dot-dashed line. As a result of the
opposite trends, Jura's formula is coincidentally correct to within a
factor of $\sim$2 over a large range of mass-loss rate. The horizontal
lines are added to guide the eye and represent the median \Mdot\ ratio
in the range 10$^{-8}$ to 10$^{-4}$ \Myr\ (dotted line) and
10$^{-7}$ to 10$^{-4}$ \Myr\ (thin solid line).
\label{fig:Jura}}
\end{figure}
%%%%%%%%%%%%%%%%%%%%%%%%%%%%%%%%%%%%%%%%%%%%%%%%%%%%%%%%%%%%%%%%%%%%%%%%%%%%%

\end{itemize}

\subsection{             Photometric Quantities                         }

The dusty envelope absorbs the stellar radiation and reradiates it at longer

wavelengths, thus making the infrared emission the most important part of the
SED for model testing. The largest catalog of infrared observations is compiled
by Gezari {\em et al.} (1993). Individual fluxes may often be more accurate
than 10\%, but due to the inhomogeneous nature of the catalog, the mean overall
accuracy is probably lower. The wavelength coverage greatly varies among the
sources and often is based only on the IRAS catalog. Recent large scale
sensitive digital surveys (e.g. infrared 2MASS, Skrutskie {\em et al.} 1997;
optical SDSS, York {\em et al.} 2000) are bound to significantly improve the
availability of accurate multi-wavelength photometry.

Given the photometric data accurate to within 10\%, the bolometric flux could
be determined with the same accuracy, at least in principle. In practice,
however, the wavelength coverage can be sparse and this shortcoming can lead to
severe errors, unless the shape of SED is known a priori. Another difficulty is
the variability of AGB stars which can also contribute significantly to
bolometric flux errors. Given all these uncertainties, the bolometric flux can
be determined to better than 20-30\% only for a small number ($\la 100$) of
well observed stars.

\subsection{                   Distance                                }

Several methods are employed to estimate distances to AGB stars. The simplest
one assumes that all AGB stars have luminosity of 10$^4$ \Lo, and determines
distance using bolometric flux. Apart from a bias in this estimate (the median
AGB luminosity is at least a factor of 2 smaller, see e.g. Habing {\em et al.}
1985, and Knauer, Ivezi\'c \& Knapp 2001), its intrinsic accuracy cannot be
better than about 20-30\% due to the finite width of the AGB luminosity
function (Jackson, Ivezi\'c \& Knapp 2002). Other methods that have been
frequently used to estimate distance include period-luminosity relations
(e.g., Whitelock, Marang \& Feast 2000), assumption that the absolute
K-band magnitude is the same for all stars, and kinematic estimates
based on radial systemic velocities. The distance errors associated with
these methods are hard to characterize and sometimes do not even have a
Gaussian distribution (e.g. kinematic distances); they may be more
accurate than a factor of two, but probably not better than 20-30\%.

The distance estimates to nearby AGB stars have been recently greatly improved
with the data obtained by the HIPPARCOS satellite. The accuracy of
HIPPARCOS distances varies from a star to star, but, nevertheless, there are
now hundreds of AGB stars with distance estimates better than 10\%.

\subsection{             Bolometric Luminosity                                }

The bolometric luminosity can be determined using the bolometric flux and
distance estimates. Assuming HIPPARCOS distances and stars with good
photometric coverage, the luminosity can determined to within 10-20\%. In more
typical cases, its uncertainty is closer to 50\% (e.g. Knauer, Ivezi\'c \&
Knapp 2001), and without a Hipparcos parallax it can be as large as a
factor of 2 (Jackson, Ivezi\'c \& Knapp 2002).

\label{lastpage}
\end{document}